\documentclass[aip,jap,reprint,groupedaddress,a4paper]{revtex4-1}
\usepackage{amsfonts,amsmath,bm}

\usepackage{color}

\usepackage{graphicx}

\newcommand{\rl}{\rangle\!\langle}
\newcommand{\rr}{\bm{r}}
\newcommand{\kk}{\bm{k}}
\newcommand{\qq}{\bm{q}}

\newcommand{\epol}{\hat{\bm{\mathcal{E}}}}

\begin{document}

\title{Intraband absorption in finite, inhomogeneous quantum dot
  stacks for intermediate band solar cells: limitations and
  optimization}  

\author{Igor Bragar}
\author{Pawe{\l} Machnikowski} 
\email{Pawel.Machnikowski@pwr.wroc.pl}
\affiliation{Institute of Physics, Wroc{\l}aw University of
Technology, 50-370 Wroc{\l}aw, Poland}

\begin{abstract}
We present a theoretical analysis of intraband optical transitions from the
intermediate pseudo-band of confined states to the conduction
band in a finite, inhomogeneous stack of self-assembled semiconductor
quantum dots. The chain is modeled with an effective Hamiltonian
including nearest-neighbor tunnel couplings and the
absorption under illumination with both coherent (laser)
and thermal radiation is discussed. We show that the absorption
spectrum already for a few 
coupled dots differs from that of a single dot and develops a structure
with additional maxima at higher energies. We
find out that this leads to an 
enhancement of the overall transition rate under solar
illumination by up to several per cent which grows with the number of
QDs but saturates already for a few QDs in the chain. The decisive
role of the strength of  
inter-dot coupling for the stability of this enhancement 
against QD stack inhomogeneity and temperature is revealed.
\end{abstract}


\maketitle

One of the ways to improve the efficiency of solar cells is to
introduce an intermediate band in the energy spectrum of a
photovoltaic structure \cite{luque97,luque12}. In this way, electrons can be
sequentially promoted from the valence band to the intermediate band
and then to the conduction band by absorbing photons with energies
below the band gap which are not converted into useful electrochemical
energy in a standard structure.  As an implementation of this concept,
a stack of quantum dots (QDs) in the intrinsic region of a p-i-n
junction solar cell has been proposed \cite{aroutiounian01}. This idea
has indeed gained some experimental support in recent years
\cite{laghumavarapu_APL_07, hubbard_APL_08, oshima08, okada09,
okada11, blokhin09, guimard10, shang_APA_11, shang_APL_11}.
Quantum-dot-embedded p-i-n solar cells show higher quantum efficiency
in near infrared range but their overall efficiency still is lower
than the efficiency of similar devices without QDs
\cite{laghumavarapu_APL_07, hubbard_APL_08, oshima08, okada09,
okada11, blokhin09, guimard10, shang_APA_11, shang_APL_11}.

On the theory side, models involving a single QD were formulated to
describe the kinetics of transitions from and into the intermediate
levels \cite{luque10,luque11}.
On the other hand, modeling of the electron states and
optical absorption in chains and arrays of QDs has been mostly limited
to infinite, periodic superlattices of identical dots 
\cite{shao07,tomic08, tomic_PRB_2010, klos09, klos10, deng11}.  
As we have shown
recently \cite{bragar11}, enhanced absorption can appear also in
finite chains of non-identical QDs but it is suppressed if the
inhomogeneity of the QD chain (leading to non-identical electron
ground state energies in the individual QDs) becomes too large.  Since
the actual QD chains are always finite (usually built of several to a
few tens of QDs) \cite{laghumavarapu_APL_07, hubbard_APL_08, oshima08,
okada09, okada11, blokhin09, guimard10, shang_APA_11, shang_APL_11}
and unavoidably inhomogeneous it is of large practical importance for
the optimal design of intermediate band photovoltaic devices to extend
the theoretical analysis to such more realistic structures.

In this paper, we study the intraband optical absorption associated
with the electron transition from the states confined in a finite
stack of quantum dots to the conduction band. 
(Fig.~\ref{fig:sketch})
We propose a relatively simple
and computation-effective model which, however,  
includes all the essential features of the system, in particular the
inhomogeneity of the energetic parameters of the dots forming the
stack  
and the coupling between them. We show that the enhanced absorption
features appear already for a few dots and lead to enhanced transition
rate from the pseudo-band of confined states to the bulk continuum. 
While this effect is relatively stable with respect to inhomogeneity,
it turns out to be associated with the transitions  
from the ground state of the stack and is washed out as soon as the
temperature becomes comparable to the width of the pseudo-band of
confined states.  
Our modeling results suggest that this detrimental temperature effect
can be to a large extent overcome by increasing the tunnel coupling
between the dots. 

The paper is organized as follows. Sec.~\ref{sec:model} defines the
model of the system. In Sec.~\ref{sec:absorption}, we briefly discuss
the theoretical description of intraband transitions in the cases of
coherent and thermal radiation. Next, in Sec.~\ref{sec:results}, we
present the results of our calculations. Sec.~\ref{sec:conclusions}
concludes the paper.

\section{Model}
\label{sec:model}

\begin{figure}[tb]
\begin{center}
\includegraphics[width=85mm]{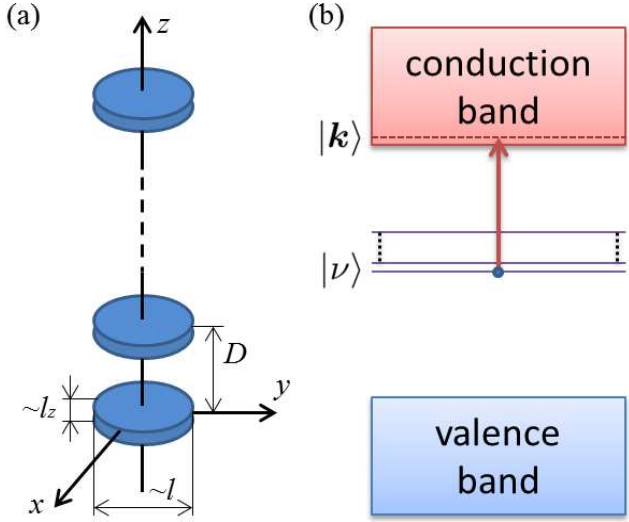}
\end{center}
\caption{\label{fig:sketch}
 (a) Sketch of a QD chain,
 (b) energy diagram of a QD chain
 with an electron transition from
 the bound state $| \nu \rangle$ 
 in the intermediate band to the state $|\bm{k}\rangle$ in the conduction band.
}
\end{figure}

We are interested in the transitions from a single-electron bound
state $|\nu\rangle$ (with a wave function $\Psi_{\nu}(\rr)$ and energy $E_{\nu}$) to a
continuum state $|\bm{k}\rangle$ (with a wave function
$\Psi_{\kk}(\rr)$ and energy $E_{\kk}=\hbar^{2}\kk^{2}/(2m^{*})$) (Fig.~\ref{fig:sketch}).  

We model the confined electron states $|\nu\rangle$ as 
superpositions of the ground states $|n\rangle$ confined in the
individual dots (where $n$ numbers the dots).  
For simplicity, we assume that each of these single dot states has an
identical wave function,  
\begin{displaymath}
\psi_{n}(\rr)=\psi_{0}(\rr-\bm{R}_{n}), 
\end{displaymath}
where $\bm{R}_{n}$ is the position of the $n$th dot 
(we assume that the dots are stacked along the growth direction $z$). 
The ground state electron energies in the dots, $\epsilon_{n}$, may differ. 
The states $|n\rangle$ are coupled by nearest neighbor couplings. 
The eigenstates $|\nu\rangle$ and the corresponding energies $E_{\nu}$
are thus obtained as the eigenstates  
of the effective chain Hamiltonian (assuming a single confined state in each dot) \cite{bragar11},
\begin{equation}\label{ham}
H=\sum_{n=0}^{N-1}\epsilon_{n}|n\rl n|
+t\sum_{n=0}^{N-2} (|n\rl n+1| +\mathrm{h.c.}),
\end{equation}
where $t$ is the coupling constant.
This coupling constant is determined by the barrier 
between the neighboring QDs.
The height of the barrier depends on the band edge mismatch 
between the QDs and on the host materials 
whereas the barrier width is set in the process of growing 
of the QD stack.
Since the stacks of self-organized QDs are produced using 
molecular beam 
epitaxy \cite{laghumavarapu_APL_07,oshima08,okada11,blokhin09,shang_APA_11} or
metal organic chemical vapor deposition  \cite{hubbard_APL_08,guimard10}
the barrier width (i.e. inter-dot
distance $D$)
is controlled with a high precision up to a single monolayer, so the coupling 
constant $t$
can be assumed to be the same for all pairs of neighboring QDs.
We assume the overlap between the
wave functions localized in different dots to be negligible, so
that $\langle n|n'\rangle=\delta_{nn'}$. The 
inhomogeneity of the QD stack is taken into account by choosing the
energies $\epsilon_{n}$ from the Gaussian distribution with the mean $\overline{E}$
and variance $\sigma^{2}$. 

We assume that the wave function for the electron in the $n$th dot has
the Gaussian form,
\begin{displaymath}
\psi_{n}(\rr)=
\frac{1}{\pi^{3/4}ll_{z}^{1/2}}
e^{-\frac{1}{2}\left[ \frac{x^{2}+y^{2}}{l^{2}}
+\frac{(z-z_{n})^{2}}{l_{z}^2}\right] },
\end{displaymath}
where $z_{n}=nD$ is the position of the $n$th dot and $l,l_{z}$ are the
extensions of the wave function in the $xy$ plane and along $z$,
respectively. 
Our choice to use the same wave function for all
QDs which have not necessarily the same ground energy levels can be argued as follows.
Using the model of quantum harmonic oscillator we can estimate 
that small differences 
of the confined energy levels in a QD (of the order of a few meV)
correspond to very small changes
of the parameters of the wave function (of the order of a few percent),
so we can approximate wave 
function of each QD by a Gaussian function with constant parameters $l$ and $l_z$.
On the other hand, when the differences of the QD confined level energies are larger 
strong 
localization of an electron on the QD with the lowest energy level occurs, which means
that the exact form of the wave functions (i.e. knowledge of the precise values of parameters) 
of other QDs become irrelevant,
so that in this case we also can use the same parameters
$l$ and $l_z$ for all QDs of the chain.

For the bulk electron states, we assume plane waves
\cite{magnusdottir02} orthogonalized to 
the localized states, as previously proposed for calculating carrier capture rates 
\cite{ferreira99, schneider_PHYSRB_01, nielsen04}. These  states are
labeled by the wave  vector $\kk$ 
describing the plane wave far away from the QD structure. Thus, we
have
\begin{displaymath}
\Psi_{\kk}(\rr)=N_{\kk}\left[ \frac{1}{\sqrt{V}}e^{i\kk\cdot\rr}
-\sum_{n}\gamma_{\kk n} \psi_{n}(\rr)  \right],
\end{displaymath}
where $N_{\kk}$ is the appropriate normalization constant, we assume
normalization in a box of volume $V$ with periodic boundary conditions,
and the orthogonalization coefficients $\gamma_{\kk n}$ are given by
\begin{displaymath}
\gamma_{\kk n}=\frac{1}{\sqrt{V}}
\int d^{3}r \psi_{n}^{*}(\rr)e^{i\kk\cdot\rr}
=e^{ik_{z}z_{n}}\gamma_{\kk 0},
\end{displaymath}
where 
\begin{displaymath}
\gamma_{\kk 0} = 
\frac{ll_{z}^{1/2}\pi^{3/4}2^{3/2}}{\sqrt{V}}
e^{-\frac{1}{2}\left[
l^{2}(k_{x}^{2}+k_{y}^{2})+l_{z}^{2} k_{z}^{2}
 \right] }.
\end{displaymath}

The coupling of carriers to the  incident light is described by the
dipole Hamiltonian
\begin{equation}
\label{Hint}
H_{\mathrm{int}}=e\rr\cdot \bm{\mathcal{E}}(\rr), 
\end{equation}
where  $e$ is the elementary charge and 
$\mathcal{E}$ is the electric field.
We will consider two cases: 
A monochromatic laser light will be described as a classical plane
wave field
\begin{equation}
\label{E-clas}
\bm{\mathcal{E}}^{(\mathrm{cl})}(\rr)=
\frac{1}{\varepsilon_{0}\varepsilon_{\mathrm{\infty}}}
\epol \mathcal{E}_{0}\cos(\qq\cdot\rr-\omega_{q} t),
\end{equation}
where $\varepsilon_{0}$ is the vacuum permittivity,
$\varepsilon_{\mathrm{\infty}}$ is the high-frequency dielectric
constant of the semiconductor, 
$\mathcal{E}_{0}$ is the amplitude of the electric field of the
electromagnetic wave, $\epol$ is a unit vector defining its
polarization, $\qq$ is its wave vector (inside the dielectric 
medium), and  $\omega_{q}=cq/n_{\mathrm{r}}$ is its
frequency, where $n_{\mathrm{r}}$ is the refractive index of the
semiconductor.  On the other hand, for thermal radiation,
corresponding to the natural working conditions of a solar cell, the
field is 
\begin{equation}
\label{E-quant}
\bm{\mathcal{E}}^{(\mathrm{th})}(\rr)=
{\sum_{\qq}}' 
\sqrt{\frac{\hbar \omega_{q}}{2\varepsilon_{0}\varepsilon_{\infty}V}}
\hat{\bm{\mathcal{E}}} a_{\qq} e^{i\qq\cdot\rr} +\mathrm{h.c.},
\end{equation}
where $a_{\qq}$ is the annihilation operator for a photon with the
wave vector $\qq$, $V$ is the formal normalization volume, and we 
take into account that the incident solar
radiation is propagating into a specific direction, hence its wave
vectors are distributed over a very small solid angle around its
direction of propagation $\hat{\bm{q}}$ (which is represented by the
prime at the 
summation sign). For more flexibility of the modeling, we
assume also that the radiation is polarized (the effects of
unpolarized radiation can be modeled by averaging over the directions
of polarization).

\section{Light absorption by a QD chain}
\label{sec:absorption}

In the description of light induced transitions from the confined
states to the extended states we assume that the occupation of the
latter is negligible, which in a solar cell corresponds to assuming efficient
carrier collection. 

In the case of classical (coherent) monochromatic light with frequency
$\omega$, propagation direction $\hat{\qq}$, and polarization
$\epol$, the transition rate 
from a state $|\nu\rangle$ to the 
continuum of extended states is obtained in the usual way from the
Fermi golden rule \cite{messiah66} using the interaction Hamiltonian
\eqref{ham} with the field given by Eq.~\eqref{E-clas},
\begin{eqnarray*}
\lefteqn{\alpha_{\nu}(\omega,\hat{\qq}, \epol) =}\\ 
&&\frac{2\pi}{\hbar^{2}}
\left(
  \frac{e\mathcal{E}_{0}}{2\varepsilon_{0}\varepsilon_{\mathrm{\infty}}} 
\right)^{2}
\sum_{\kk}
\left| \left\langle 
\Psi_{\kk} |\rr\cdot \epol e^{i\qq\cdot\rr} | \Psi_{\nu}
\right\rangle \right|^{2} \delta\left( \omega-\omega_{\kk\nu} \right).
\end{eqnarray*}
where $\omega_{\kk\nu}=(E_{\kk}-E_{\nu})/\hbar$.
This can be written in the form 
\begin{equation}
\alpha_{\nu}(\omega,\qq, \epol)=
\frac{\pi e^{2}\mathcal{U}}{
\hbar^{2}\varepsilon_{0} \varepsilon_{\mathrm{\infty}}}
\sum_{\kk}|\alpha_{\nu \kk}(\qq)|^{2}\delta(\omega-\omega_{\kk\nu}),
\label{abs}
\end{equation}
where
$\mathcal{U}=\varepsilon_{0}\varepsilon_{\mathrm{\infty}}E_{0}^{2}/2$
is the energy density of the electromagnetic wave and
\begin{equation}
\alpha_{\nu \kk}(\qq)=
\int d^{3}r \Psi_{\kk}^{*}(\rr) e^{i\qq\cdot\rr}\rr\cdot\epol
\Psi_{\nu}(\rr).
\label{amplitude}
\end{equation}

In the case of illumination by broad band thermal radiation, 
we first use the Hamiltonian \eqref{Hint} with the quantum field given
by Eq.~\eqref{E-quant} to calculate the Fermi golden rule probability
of absorption of a photon with a wave vector $\qq$,
\begin{displaymath}
\gamma_{\nu}(\qq) = \frac{2\pi}{\hbar} \sum_{\kk} 
\frac{e^{2}\hbar\omega_{\qq}}{2\varepsilon_{0}\varepsilon_{\infty}}
\left| \alpha_{\nu\kk}(\qq) \right|^{2} n_{q}
\delta \left(\omega_{\qq}-\omega_{\kk\nu} \right),
\end{displaymath}
where 
$n_{q}$ is the Bose distribution of photon occupations at the
temperature of solar black body radiation and the transition
matrix elements $\alpha_{\nu\kk}(\qq)$ are given by
Eq.~\eqref{amplitude}. We note that for 
radiation propagating in a fixed direction $\hat{\bm{q}}$, the
overlap integral $\alpha_{\nu \kk}(\qq)$ actually depends only on the
frequency,
$\alpha_{\nu \kk}(\qq)=
\alpha_{\nu\kk}(n\omega_{q}\hat{\bm{q}}/c)$. 
Then, the total photon absorption rate for an initial confined state
$|\nu\rangle$ per unit frequency interval is  
\begin{eqnarray*}
\lefteqn{\beta_{\nu}(\omega, \hat{\qq}, \epol)  =  
{\sum_{\qq}}'\delta(\omega-\omega_{\qq})\gamma_{\nu}(\qq)}\\
&& = \sum_{\kk} \left| 
\alpha_{\nu\kk}\left(\frac{n\omega_{\qq}}{c}\hat{\bm{q}} \right)
 \right|^{2} \frac{\pi e^{2}}{\hbar
   \varepsilon_{0}\varepsilon_{\infty}}
\delta\left( \omega_{\qq}-\omega_{\kk\nu} \right) \\
&&\quad \times
\frac{1}{V} {\sum_{\qq}}' \hbar\omega_{\qq}n_{q}
\delta(\omega-\omega_{\qq}).
\end{eqnarray*}
The final sum in this expression is the spectral distribution of the
energy density of radiation $u(\omega)$ which, for solar light, is
approximately described by the Planck law. Hence, the absorption rate
under illumination by thermal radiation can be written in the form
analogous to Eq.~\eqref{abs},
\begin{equation}
\beta_{\nu}(\omega, \hat{\qq}, \epol)=
\frac{\pi e^{2}u(\omega)}{
\hbar\varepsilon_{0} \varepsilon_{\mathrm{\infty}}}
\sum_{\kk}|\alpha_{\nu \kk}(\qq)|^{2}\delta(\omega-\omega_{\kk\nu}).
\label{beta}
\end{equation}

Eqs.~\eqref{abs} and \eqref{beta}, along with the effective chain
Hamiltonian \eqref{ham}, are the basis for
numerical calculations the results of which are presented in the next section.

\section{Results}
\label{sec:results}

In this section, we present results of numerical calculations of the
intraband absorption in QD chains using the approach described in
Sec.~\ref{sec:model} and Sec.~\ref{sec:absorption}.

In all our simulations we focus on an InAs/GaAs structure (not optimal
for a solar cell \cite{luque97} but important for the current
laboratory scale investigations \cite{luque12}). We assume the wave
function confinement sizes  
$l=4.5$ nm, $l_{z}=1.0$ nm (see Ref.~\onlinecite{bragar11} for the
discussion of the dependence on these parameters), $\overline{E}=-250$
meV.
Based on earlier $\bm{k} \cdot \bm{p}$ calculations
\cite{gawarecki_PRB_2010}, the tunnel coupling $t$ is given by the
formula 
\begin{displaymath}
\ln \frac{t}{t_{0}} = - \mathcal{K}D,
\end{displaymath}  
where $\mathcal{K}=0.59 ~\text{nm}^{-1}$, $t_0=0.79$ eV ($t=-3.9$ and
$-41.3$ meV for $D=9$ and $5$ nm, respectively). 

\begin{figure}[tb]
\begin{center}
\includegraphics[width=85mm]{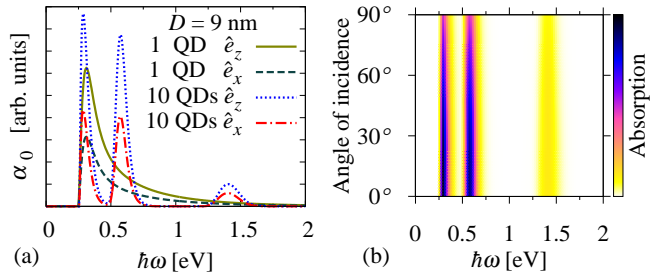}
\end{center}
\caption{\label{fig:length}(a) Low temperature intraband absorption
 spectra for a chain of identical dots illuminated by 
 coherent monochromatic
 unpolarized light incident along  
 and perpendicular to the stacking direction $z$ (as indicated by
 $\hat{e}_z$ and $\hat{e}_x$, respectively).  
 (b) The dependence of the spectrum on the angle of incidence for 10~QDs.}
\end{figure}

In Fig.~\ref{fig:length}(a), we show the intraband absorption spectra
of short chains of identical QDs compared to the spectrum of a single
dot  
for light incident along the chain (in the $z$ direction) and perpendicular to the chain.  
Both spectra are calculated assuming that only the lowest confined
state is considerably occupied (which corresponds to low temperatures)  
and for unpolarized light (for perpendicular incidence, the contribution from light polarized along $z$ is small). 
In both cases,
the absorption spectrum for a single dot shows only a single  maximum
with a long tail on the high energy  side but already for a few dots  
it develops a series of additional maxima. As can be seen in
Fig.~\ref{fig:length}(b), the spectra for a chain of 10 QDs smoothly evolve between the
two limiting geometries as the incidence angle is tilted from normal to in-plane incidence.

\begin{figure}[tb]
\begin{center}
\includegraphics[width=85mm]{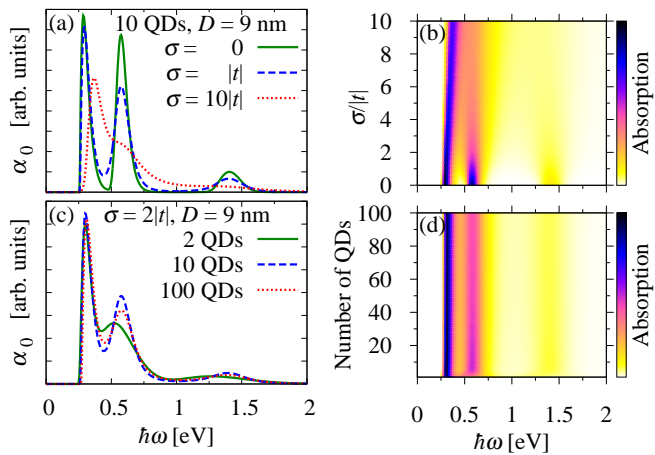}
\end{center}
\caption{\label{fig:inhom}Low temperature intraband absorption
 spectrum  for a chain of non-identical dots  
 illuminated by
 coherent monochromatic
 unpolarized light incident along the stacking direction $z$.
 (a,b) As a function of the degree of inhomogeneity; (c,d) as a
 function of the chain length.  
 Each spectrum is an average of 1000 realization with randomly chosen
 values of $\epsilon_{n}$.} 
\end{figure}

The additional absorption peaks that can be seen in
Fig.~\ref{fig:length} result from interference effects which are due
to delocalization  
of the electron state and lead to preferred transitions to states with
$k_{z}=2\pi j/D$, where $j$ is an integer \cite{bragar11}.  
Disorder, which in our case has the form of inhomogeneity of the
``on-site'' energies $\epsilon_{n}$,  
destroys the coherently delocalized electron states and leads to
localization. As a result, the additional peaks are suppressed  
and the absorption spectrum becomes more similar to that of a single
dot (see Fig.~\ref{fig:inhom}(a,b)). 
As we show in Fig.~\ref{fig:inhom}(c,d), while the spectra for QD
stacks differ considerably from those for a single dot, the extension
of the stack above $10$ dots has little effect.
Note that the effect of interference of transition amplitudes depends
to some extent on whether their magnitudes are equal or not. Hence, if
one allows different geometries of wave functions in non-identical QDs
then additional suppression of the interference-related additional
peaks may occur. However, this effect is expected to be small compared
to the suppression due to localization.

\begin{figure}[!t]
\begin{center}
\includegraphics[width=85mm]{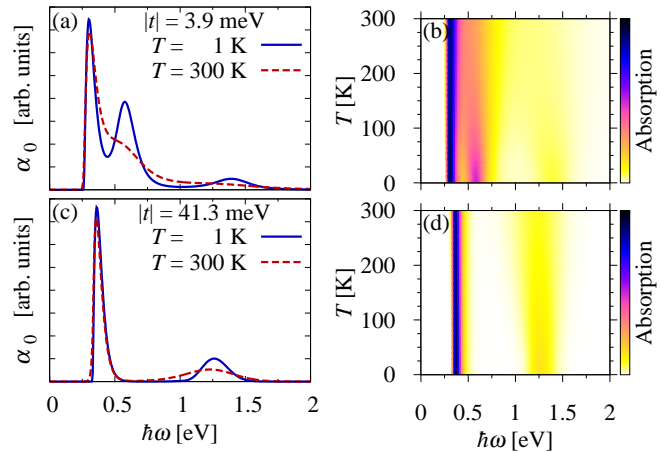}
\end{center}
\caption{\label{fig:temp}Intraband absorption spectrum for a chain of
 10 non-identical dots with $\sigma=7.8$~meV,
 illuminated by 
 coherent monochromatic
 unpolarized light incident along the stacking direction
 $z$ as a function of temperature  
 for two values of the inter-dot coupling: (a,b) $t=-3.9$ meV 
 (corresponding to $D=9$ nm), (c,d) $t=-41.3$ meV (corresponding to
 $D=5$ nm).}  
\end{figure}

For a realistic device, absorption at elevated temperatures is
relevant. At non-zero temperatures, the occupation of excited states  
in the QD-related pseudo-band is non-negligible, which leads to a
reconstruction of the spectrum, as shown in Fig.~\ref{fig:temp}.  
For chains with $|t| < kT$, the structure of the spectrum is washed
away with increasing temperature (Fig.~\ref{fig:temp}(a,b)). 
However, as can be seen from Fig.~\ref{fig:temp}(c,d) strong coupling
broadens the pseudo-band beyond the thermal energies and stabilizes the
chain absorption against temperature.  

\begin{figure}[!t]
\begin{center}
\includegraphics[width=85mm]{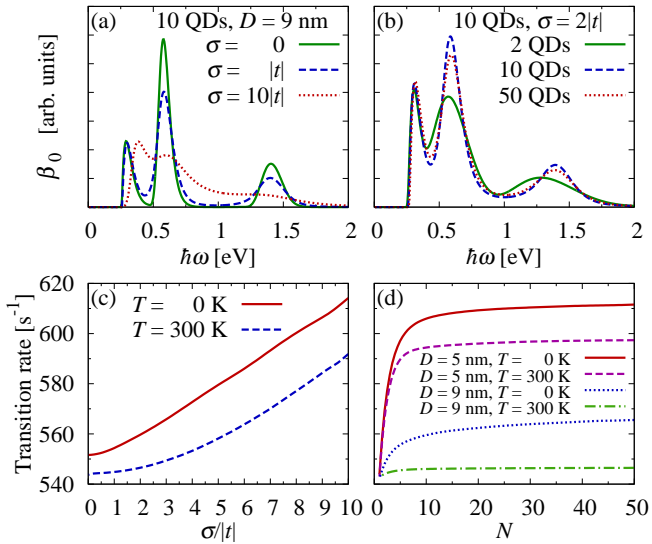}
\end{center}
\caption{\label{fig:beta}
(a,b) Electron intraband transition rates under illumination by one
sun thermal radiation
as a function of the degree of inhomogeneity (a) and  
as a function of the number of QDs in the chain (b). Each spectrum is
an average of 1000 realization with randomly chosen values of
$\epsilon_{n}$. 
(c,d) Integrated single-electron transition rate under illumination by
one sun thermal radiation 
as a function of the degree of inhomogeneity (c) and as a function of
the number of QDs in the chain(d).}
\end{figure}

In order to study the properties of the device under realistic
working conditions we calculate also  the electron intraband transition
rate $\beta_{0}$ (from the ground state) for a chain of non-identical QDs 
illuminated by black body radiation. The energy
density of radiation with $T = 5777$~K was normalized to the incident 
flux of 1~kW/m$^2$ (one sun). 
In Fig.~\ref{fig:beta}(a,b) we show the spectral distribution of the
transition rates as a 
function of the energy of the absorbed photon. Comparison with the
absorption spectra shown in Fig.~\ref{fig:length} shows that the
qualitative form of these two spectral characteristics is similar but quantitatively
the transition rate at higher photon energies is enhanced relative to
that at lower energies. This is due to the very quickly growing
spectral densities of the thermal photon field at higher energies (according to the
the Planck law) which enhances the role of higher energy photons (and,
in consequence, the overall absorption of energy in the corresponding
spectral range). One should notice that inhomogeneity (higher values
of $\sigma$) leads to a less structured spectrum with non-vanishing
absorption over all the spectral range.

In fig Fig.~\ref{fig:beta}(c), we show the total transition rate,
integrated over photon energies. The result at $300$~K includes the
thermal distribution of the initial states of the electron.
As one can see, the total rate increases as the inhomogeneity grows.
This increase is due to the increasing probability that the electron
will localize in a dot with much lower ``on-site'' energy  
as the QD energy distribution broadens. This enhances the transition
rate because of the growing spectral density of radiation for higher
energy differences between the initial and final states.
As a function of the chain length (Fig.~\ref{fig:beta}(d)), the total
transition rate increases slightly (up to several \%) for strongly coupled
chains at low temperatures
and then saturates for $N \gtrsim 10$. This
increase is mostly due to high-frequency contributions corresponding
to the enhanced high-energy absorption features in a QD stack
(cf. Fig.~\ref{fig:length}).  
One can notice in Fig.~\ref{fig:beta}(d) that the gain in the total
transition rate for strongly coupled dots not only is much larger than
for weaker coupling but it is also much less sensitive to temperature.

\section{Conclusions}
\label{sec:conclusions}

We have proposed a model which describes electron states
and intraband absorption spectra which correspond to the transitions from
the QD pseudo-band to the conduction band.  We have studied the
absorption spectra as a function of the degree of inhomogeneity, the
number of QDs in the chain, the inter-dot separation, the temperature
and the illumination conditions.  We have shown that the absorption
spectrum even of short QD chains (a few QDs) is dominated by
interference effects that leads to the appearance of additional
absorption maxima and does not considerably evolve further if
the chain length exceeds approximately 10 dots.  The overall
transition rate under illumination with black body radiation increases
by up to several per cent with the number of QDs but also saturates
already for a few QDs in the chain. The structured absorption spectrum
persists when the inhomogeneity of the QD transition energies is
increased up to the values comparable with the magnitude of the
inter-dot coupling. Strong coupling between the dots is essential for
maintaining the chain absorption features up to high temperatures.  On
practical side, our results show that already a stack of a few QDs
manifests the absorption features characteristic of QD chains and that
these features are stable against temperatures and energy
inhomogeneities if the dots are sufficiently strongly coupled.
However, a QD chain shows only a slightly increases absorption as
compared to a single dot and the increase of absorption is dominated
by high-energy photons, which may lead to competition with the
interband absorption. Moreover, a trade-off has to be found  between
the absorption enhancement in strongly coupled chains and the need to
reduce tunneling in order to avoid carrier escape from the
intermediate bands \cite{antolin10}. 

\acknowledgments

This work was funded from the TEAM programme of the 
Foundation for Polish Science, co-financed by the European Regional
Development Fund.


\end{document}